\documentclass[reprint,amsmath,amssymb,nofootinbib,aps]{revtex4-1}
\usepackage{bm}
\usepackage{caption}
\usepackage{color}
\usepackage{epstopdf}
\usepackage{graphicx}
\usepackage{lmodern}
\usepackage{mathtools}
\usepackage{multirow}
\usepackage{makecell}
\usepackage{subfigure}
\usepackage{textcomp}
\usepackage[normalem]{ulem}
\usepackage{url}
\usepackage{booktabs}

\begin{document}

\title{Global constraint on the magnitude of anomalous chiral effects in heavy-ion collisions}

\author{Wen-Ya Wu$^{a,b}$}
\author{Qi-Ye Shou$^{a,b}$}
\author{Panos Christakoglou$^{c}$}
\author{Prottay Das$^{d}$}
\author{Md. Rihan Haque$^{e}$}
\author{Guo-Liang Ma$^{a,b}$}
\author{Yu-Gang Ma$^{a,b}$}
\author{Bedangadas Mohanty$^{d}$}
\author{Chun-Zheng Wang$^{a,b}$}
\author{Song Zhang$^{a,b}$}
\author{Jie Zhao$^{a,b}$}
\affiliation{$^a$Key Laboratory of Nuclear Physics and Ion-beam Application (MOE), Institute of Modern Physics, Fudan University, Shanghai 200433, China}
\affiliation{$^b$Shanghai Research Center for Theoretical Nuclear Physics, NSFC and Fudan University, Shanghai 200438, China}
\affiliation{$^c$Nikhef, Nationaal Instituut voor Subatomaire Fysica, Amsterdam, Netherlands}
\affiliation{$^d$National Institute of Science Education and Research, Homi Bhabha National Institute, Jatni, India}
\affiliation{$^e$Warsaw University of Technology, Warsaw, Poland}

\begin{abstract}
When searching for anomalous chiral effects in heavy-ion collisions, one of the most crucial points is the relationship between the signal and the background. 
In this letter, we present a simulation in a modified blast wave model at LHC energy, which can simultaneously characterize the majority of measurable quantities, in particular, the chiral magnetic effect (CME) and the chiral magnetic wave (CMW) observables. 
Such a universal description, for the first time, naturally and quantitatively unifies the CME and the CMW studies and brings to light the connection with the local charge conservation (LCC) background. Moreover, a simple phenomenological approach is performed to introduce the signals, aiming at quantifying the maximum allowable strength of the signals within experimental precision. Such a constraint provides a novel perspective to understand the experimental data and sheds new light on the study of anomalous chiral effects as well as charge dependent correlations.
\end{abstract}

\maketitle

$\it{Introduction. -}$
Collisions between heavy ions at ultra-relativistic energies have been extensively used in the last decades to study the transition to a deconfined state of matter, the quark gluon plasma. This transition, according to quantum chromodynamics (QCD) calculations on the lattice, is expected to take place at energy densities and temperatures which are accessible in the laboratories such as the Relativistic Heavy Ion Collider (RHIC) and the Large Hadron Collider (LHC). In addition, such collisions provide the unique opportunity to test novel QCD phenomena that are directly connected to the rich structure of the vacuum of the theory~\cite{Lee:1973iz,Lee:1974ma}. These phenomena are associated with transitions that lead to chirality imbalance and consequently to $\cal P$ (parity) and/or $\cal CP$ (charge-parity) violating effects in strong interactions~\cite{Morley:1983wr,Kharzeev:1998kz,
  Kharzeev:1999cz,Kharzeev:2015kna,Kharzeev:2007tn,Kharzeev:2007jp,Fukushima:2008xe}. Theoretical studies highlighted that in the presence of an external strong magnetic field, like the one generated at the initial stages of a heavy ion collision~\cite{Skokov:2009qp,Bzdak:2011yy,Deng:2012pc}, such transitions can lead to the development of macroscopic phenomena such as the Chiral Magnetic Effect (CME)~\cite{Voloshin:2004vk, Kharzeev:2004ey, Fukushima:2008xe, Wang:2018ygc} and the Chiral Magnetic Wave (CMW)~\cite{Burnier:2011bf, Burnier:2012ae, Taghavi:2013ena, Yee:2013cya}. Both the CME and the CMW are argued to have an experimentally accessible signal, see Refs.~\cite{Kharzeev:2015znc, Hattori:2016emy, Zhao:2019hta, Liu:2020ymh, Gao:2020vbh} for the latest review.

Specifically, the CME is theorized to manifest itself in a finite electric dipole moment in the quark-gluon plasma (QGP) and develops along the direction of the magnetic field. Taking advantage of the azimuthal emission of final state hadrons, it's feasible to detect the CME-induced signal via the $\gamma$ and $\delta$ correlators~\cite{Voloshin:2004vk, Liao:2010nv}:
\begin{equation} \label{eq:1}
\gamma \equiv {\rm cos}(\phi_\alpha+\phi_\beta-2\Psi), \quad \delta \equiv {\rm cos}(\phi_\alpha-\phi_\beta),
\end{equation}
where $\phi_\alpha$ and $\phi_\beta$ are azimuthal angles of two particles of interest, and $\Psi$ is that of the reaction plane, the plane defined by the impact parameter between the two colliding nuclei and the beam axis. This measurement is usually performed with the same-sign (SS) and opposite-sign (OS) charge combinations of $\alpha$ and $\beta$, and their differences are used to explore the possible signal:
\begin{equation} \label{eq:2}
\Delta\gamma \equiv \gamma_{\rm OS}-\gamma_{\rm SS}, \quad \Delta\delta \equiv \delta_{\rm OS}-\delta_{\rm SS}.
\end{equation}
Meanwhile, the CMW is expected to create an electric quadrupole moment in the participant region, where the ``poles" (out of plane) and the ``equator" (in plane) respectively acquire additional positive or negative charges~\cite{Burnier:2011bf}. Such an effect can be probed by the charge asymmetry ($A_{\rm ch}$) dependence of elliptic flow ($v_2$) between the positively and negatively charged particles:
\begin{equation} \label{eq:3}
\Delta v_{2} \equiv v_{2}^{-} - v_{2}^{+} \simeq rA_{\rm ch},
\end{equation}
where $A_{\rm ch} \equiv (N^{+} -N^{-}) / (N^{+} +N^{-})$ with $N$ denoting the number of particles in a given event, and the slope $r$ is used to quantify the signal.

\begin{table*}
\renewcommand\arraystretch{1.6}
\captionsetup{justification=raggedright}
\setlength{\tabcolsep}{2mm}
\caption{List of the modified BW parameters for Pb-Pb collisions at $\sqrt{s_{\rm NN}}$ = 5.02 TeV.}
\begin{tabular}{c || c | c | c | c | c | c | c | c}
Centrality & 0-5\% & 5-10\% & 10-20\% & 20-30\% & 30-40\% & 40-50\% & 50-60\% & 60-70\% \\
\midrule[1.5pt]
$T_{\rm kin}$ & 111.34 &106.96 &104.78 &107.37 &111.63 &115.14 &118.14 &128.20 \\
\hline
$R_x/R_y$ & 0.956 & 0.934 & 0.905 & 0.872 & 0.845 & 0.823 & 0.807 & 0.786 \\
\hline
$\rho_0$ & 1.262 & 1.267 & 1.254 & 1.226 & 1.196 & 1.148 & 1.087 & 0.994 \\
\hline
$\rho_2$ & 0.054 & 0.063 & 0.11 & 0.135 & 0.15 & 0.145 & 0.121 & 0.115 \\
\midrule[1.5pt]
$N_{\rm ch}^{~|\eta|< 0.8}$ & 2290 & 1858 & 1334 & 904 & 608 & 369 & 222 & 117 \\
\hline
$f_{\rm LCC}$ & 0.71 & 0.62 & 0.58 & 0.56 & 0.54 & 0.48 & 0.47 & 0.46 \\
\end{tabular}
\label{tab:tab1}
\end{table*}

Over the past decade, the charge separations caused by the CME and the CMW have been carefully sought by the STAR~\cite{Abelev:2009ac,Abelev:2009ad,STAR:2013ksd,STAR:2014uiw,STAR:2019xzd,STAR:2019bjg,STAR:2021mii,STAR:2021pwb,Choudhury:2021jwd,STAR:2015wza,STAR:2022hfy,Xu:2020sln}, ALICE~\cite{Abelev:2012pa,ALICE:2015cjr,ALICE:2017sss,Acharya:2020rlz} and CMS~\cite{Sirunyan:2017tax, Sirunyan:2017quh, Khachatryan:2016got} experiments at different collision energies and systems with multiple observables. Though early data suggest some hints matching theoretical expectations, it is soon found that the background effects play a dominant role in experimental measurements. In both CME and CMW studies, for instance, the observables dramatically vary as $v_2$ changes~\cite{ALICE:2017sss, Wu:2022}, indicating significant contributions from the interplay between the way particles are being produced in pairs of oppositely charged partners, referred to as local charge conservation (LCC) and collective flow. Accounting for this background in the measurement reveals that the signal is consistent with zero within uncertainties. To understand the background and to disentangle the signal, various theoretical and phenomenological models containing different background and signal sources are developed to interpret the data, such as AVFD~\cite{Shi:2018sah,Shi:2019wzi,Christakoglou:2021nhe}, AMPT~\cite{Lin:2004en,Ma:2011uma,Ma:2014iva}. These models succeed in describing some aspects of the measurements while, unfortunatelly, fail in others. An one-size-fits-all description reasonably coinciding with most experimental observables remains incomplete.

The CME and the CMW are usually treated as two independent analyses up to now, via methods in Eq. (\ref{eq:1}, \ref{eq:2}) and (\ref{eq:3}) respectively. Even though the collectivity-convolved LCC is now clearly realized to be the background for both, no attempt has been made yet to unify the studies of the CME and the CMW, in particular, to estimate a comprehensive background in a realistic environment comparable to experimental data. 
In this letter, we present simulation results using a modified blast wave (BW) model~\cite{Retiere:2003kf,Tomasik:2008fq} at LHC energy. We will show that, for the first time, when most global observables (e.g. $p_{\rm T}$ spectrum, $v_2$, charge balance function) are tailored to describe experimental measurements (the data-driven way), the CME and CMW observables (e.g. $\Delta\gamma$, $\Delta\delta$, $A_{\rm ch}$-$v_2$ slope) can be simultaneously and naturally reproduced. Based on that, the maximum allowable strength of the signal for both CME and CMW is further deduced through a phenomenological method, which sheds new light on the search for anomalous chiral effects.

$\it{Methodology. -}$
The blast wave model is extensively used in heavy-ion collisions~\cite{Retiere:2003kf,Tomasik:2008fq}, providing a convenient and straightforward way to describe the production of particles as well as their collective motions. It generates an expanding and locally thermalized fireball, which decays into fragments and subsequently emits hadrons. The thermal equilibrium of hadrons is based on the Boltzmann distribution with kinetic freeze-out temperature $T_{\rm kin}$. The phase space distribution of the fragments is determined on the assumption that the radial expansion velocity is proportional to the distance from the centre of the system. The initial shape of the fireball is controlled by a geometry parameter $R_x/R_y$ describing the spacial asymmetry and the collectivity is determined by the radial flow parameter $\rho_0$ and the elliptic flow parameter $\rho_2$ in form of $\rho \cos(2\phi)$ with $\phi$ being the boost angle. For simplicity, higher order flow components are omitted since $v_2$ is the leading-order term and all particles are set to have pion mass.
This study is performed in Pb-Pb collisions at $\sqrt{s_{\rm NN}}$ = 5.02 TeV and the aforementioned parameters are listed in the top four rows of Tab.~\ref{tab:tab1}, with which the basic feature of $p_{\rm T}$ distribution and $v_2$ measured by ALICE experiment~\cite{ALICE:2019hno, ALICE:2018rtz} can be well reproduced. For instance, the calculated integrated $v_2$ within $p_{\rm T}<2$~GeV/$c$ and $|\eta|<0.8$ at 5-10\%, 30-40\%, 50-60\% centrality intervals are 0.035$\pm 1\cdot10^{-5}$, 0.081$\pm 1.5\cdot10^{-5}$, 0.078$\pm 3\cdot10^{-5}$, respectively, matching experimental values~\cite{ALICE:2018rtz} within 3\% relative deviations.

Of particular importance for this work is the joint treatment of the multiplicity and the balancing charge. A realistic multiplicity value is usually taken into account for the study of particle production, however, is sometimes ignored for studies of correlations. For charge dependent correlations, we argue that a comparable multiplicity is significantly important because of two reasons: (I) In the CME study, the transverse momentum conservation (TMC) plays a non-negligible role. As pointed out in Refs.~\cite{Pratt:2010zn,Bzdak:2010fd}, inclusive $\gamma$ and $\delta$ can be written into $\langle cos\phi \rangle^2 - \langle sin\phi \rangle^2 - v_2/N$ and $\langle cos\phi \rangle^2 + \langle sin\phi \rangle^2 - 1/N$ respectively, with $\langle\rangle$ being the event average. It is obvious that both $\langle sin\phi \rangle$ and $\langle cos\phi \rangle$ obey the law of large numbers (the central limit theorem as well), which manifests itself through the multiplicity. Events with high or low multiplicity (tight or loose TMC condition), therefore, naturally give rise to different $\gamma$ and $\delta$ results. (II) In the CMW study, as shown in Eq.(\ref{eq:3}), the $A_{\rm ch}$ is a key observable determining the signal. By definition, the $A_{\rm ch}$ follows a negative binomial (NBD) distribution. The larger the multiplicity is, the narrower the $A_{\rm ch}$ distribution would be. Here, in each centrality, we sample the multiplicity event by event following the NBD with mean and variance from the ALICE results~\cite{ALICE:2019hno}. It will be shown later that the $A_{\rm ch}$ distribution can  be precisely reproduced.

\begin{figure}
\centering
\includegraphics[width=0.5\linewidth]{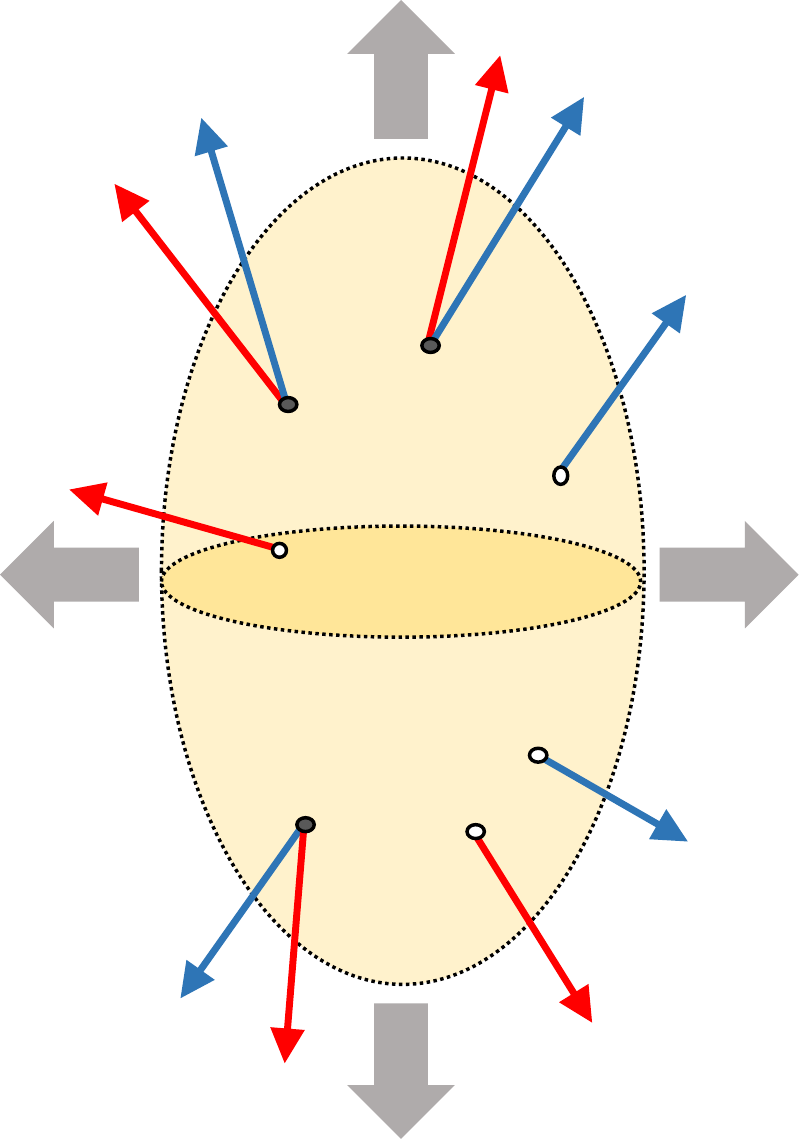}
\captionsetup{justification=raggedright}
\captionof{figure}{A schematic view of a BW event with improved treatment of the LCC. Grey arrows indicate the collective expansion. Positive and negative charges are marked in red and blue respectively. Open and full circles denote the single particle production and the pair production respectively.}
\label{fig:fig1}
\end{figure}

It is well known that, in heavy-ion collisions, for every produced particle in a specific phase space window, there should be a corresponding antiparticle with the opposite charge, which is commonly quantified by the balance function~\cite{Bass:2000az}. To introduce such a local charge dependent correlation, particles are emitted in pairs with conserved charge (one positively and one negatively charged) at some spatial points which are uniformly distributed within an ellipse, as illustrated in Fig.~\ref{fig:fig1}. The momenta of particles in a given pair are independently sampled and then boosted together so particles eventually follow a common collective velocity. For the rest of the spatial points, only one particle is generated with random charge. The percentage of points emitting pairs, $f_{\rm LCC}$, is used to parameterize the strength of LCC. We tune the value of $f_{\rm LCC}$ in each centrality to match the experimental result of charge balance function, e.g., in 30-40\%, the $f_{\rm LCC}=0.54$ gives the width of the balance function of 0.64$\pm0.014$, which is consistent with the ALICE measurement~\cite{ALICE:2013vrb,ALICE:2015nuz,ALICE:2021hjb} within 5\% relative deviations.

\begin{figure}
\centering
\includegraphics[width=\linewidth]{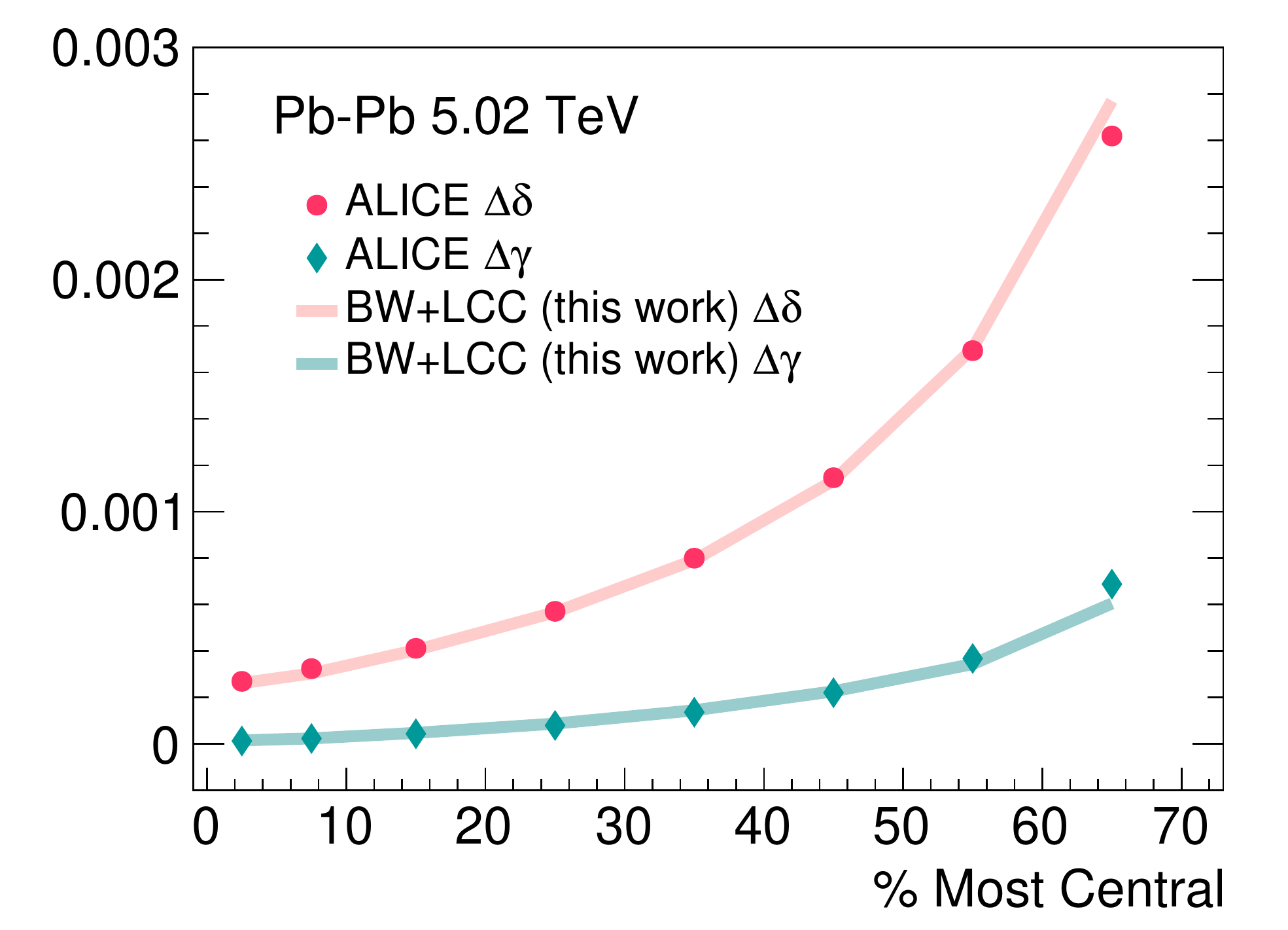}
\captionsetup{justification=raggedright}
\captionof{figure}{The CME observables $\Delta\gamma$ and $\Delta\delta$ as functions of centrality. The ALICE results are from~\cite{Acharya:2020rlz}.}
\label{fig:fig2}
\end{figure}

$\it{CME~and~CMW~observables. -}$ 
Figure~\ref{fig:fig2} presents the CME observables $\Delta\gamma$ and $\Delta\delta$ as functions of centrality on the basis of the aforementioned model configuration. It is not surprising that the ALICE measured $\Delta\delta$ can be perfectly described by the model since, as mentioned in~\cite{Acharya:2020rlz, Hori:2012kp}, $\Delta\delta$ is an equivalent form of the charge balance function. More importantly, the calculated $\Delta\gamma$ values are found to be quantitatively in line with ALICE data as well within 5\% relative deviations in central and semi-central collisions. In the most peripheral collisions, the deviation increases to 10\%, mainly owing to the fluctuation at low multiplicity. In Ref~\cite{Acharya:2020rlz}, ALICE performed a similar comparison and found that when tuning the BW model to match the measured $\Delta\delta$, the model underestimates the measured $\Delta \gamma$ by around 40\%. We would like to point out that such a discrepancy actually comes from the incomplete treatment of the LCC fraction and the multiplicity. With such an issue being properly addressed, a quantitative description of both $\Delta \delta$ and $\Delta \gamma$ can be achieved.

\begin{figure}
\centering
\includegraphics[width=\linewidth]{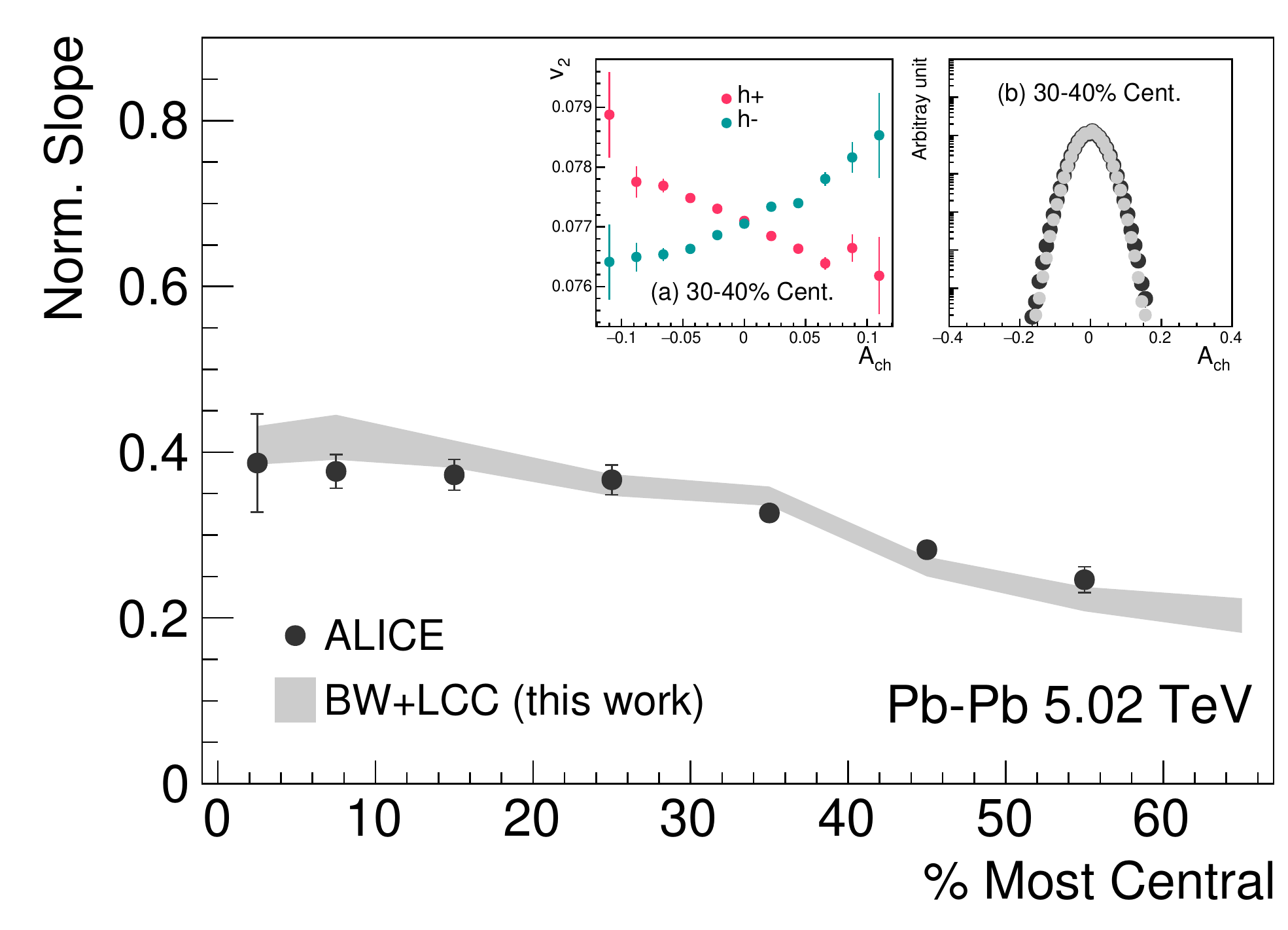}
\captionsetup{justification=raggedright}
\captionof{figure}{The CMW observable, normalized slope of $A_{\rm ch}$-$\Delta v_2$, as a function of centrality. The ALICE results are from~\cite{Wu:2022}. Embedded panel: (a) the linear dependence between $v_2^{\pm}$ and $A_{\rm ch}$, (b) $A_{\rm ch}$ distribution, in 30-40\% centrality.}
\label{fig:fig3}
\end{figure}

In addition to the CME study, the BW+LCC model has also been used for the study of CMW. As demonstrated in our early work~\cite{Wang:2021nvh}, when selecting events with a specific $A_{\rm ch}$, in practice, one preferentially applies nonuniform kinematic cuts on charged particles and such a LCC background is too ubiquitous to be eliminated. The underlying mechanism has also been clearly discussed in Refs~\cite{Bzdak:2013yla, Voloshin:2014gja, Wu:2020wem}. Nevertheless, an accurate estimation remains unexplored. Fig.~\ref{fig:fig3} shows the calculated normalized slope of $A_{\rm ch}$-$\Delta v_2$ in this model. It can be seen that the slope values match the ALICE measurement within 12\% relative deviation. The centrality dependence can be naturally described as well: the slight decrease of the slope is due to the smaller number of balancing pairs when collisions becomes more peripheral. As a detailed example, the linear dependence between $v_2^{\pm}$ and $A_{\rm ch}$ and the $A_{\rm ch}$ distribution in 30-40\% centrality are attached in the embedded panel (a) and (b) respectively. Note that the $A_{\rm ch}$ distribution is in good accord with the ALICE result~\cite{Wu:2022}, supporting our explanation in the previous section.

The set of parameters summarized in Tab.~\ref{tab:tab1} may not be the only viable configuration, however, given the fact that the small experimental uncertainty of $\Delta \gamma$ ($\sim10^{-4}$) imposes a strong constraint on the model, there is little room for those parameters to change.

Taken together, the consistency between the modified BW model and the ALICE data suggests that the current experimental measurements of anomalous chiral effects can be reasonably explained by a simple and realistic LCC background. 
Nevertheless, it does not yet rule out signals of the CME and the CMW. Even though the signals are widely believed to be small~\cite{Muller:2010jd}, from the perspective of observables, there is a possibility that, in principle, both CME and CMW can exist concurrently together with a slightly weaker LCC background. On the basis of the model with a comprehensive understanding of the background, we take one step further to determine the maximum allowable strength of signals by introducing charge separations in a phenomenological approach.

\begin{figure}
\centering
\includegraphics[width=\linewidth]{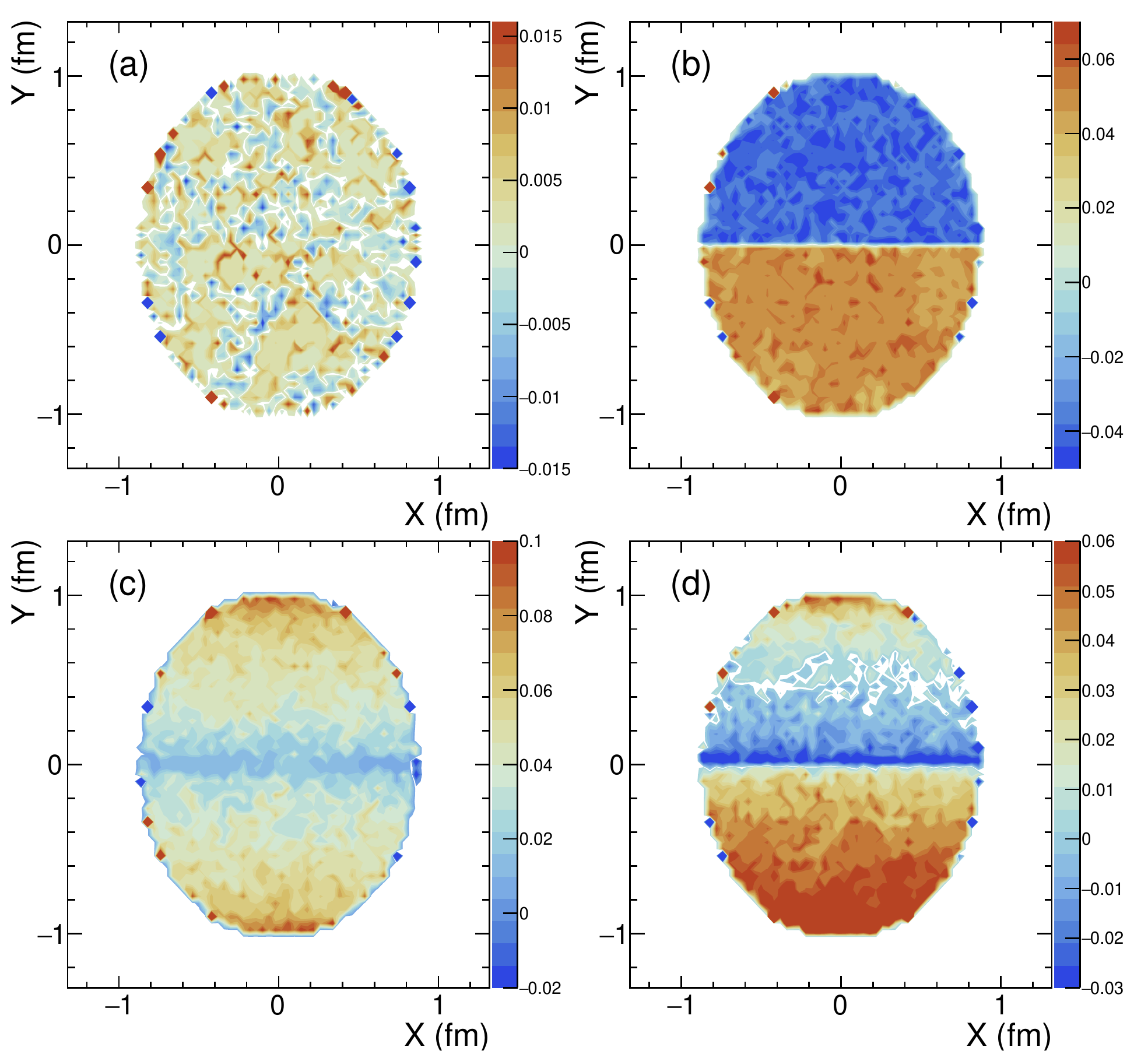}
\captionsetup{justification=raggedright}
\captionof{figure}{The net electric charge distributions in the transverse plane when the signal is (a) not imported, (b) imported to mimic the CME-induced dipole moment, (c) imported to mimic the CMW-induced quadrupole moment, (d) imported to mimic a superposed effect of both CME and CMW. Signals are intentionally enhanced here for better visualization.}
\label{fig:fig4}
\end{figure}

$\it{Constraint~on~the~Signal. -}$ 
In previous AMPT studies~\cite{Ma:2011uma,Ma:2014iva}, the $y$ component of the momentum (position) coordinate for some ``above-plane" (in-plane) particles carrying a given charge are randomly interchanged with those ``below-plane'' (out-of-plane) ones carrying the opposite charge to mimic the CME (CMW) signal. Such a straightforward operation has proven to be fairly effective. In this work, it should be emphasized that, instead of switching particles' positions or momenta, we simply interchange their charges, which is indeed equivalent to the former when the evolution is not taken into account. Specifically, only charges of those single-produced particles, denoted by white dots in Fig.~\ref{fig:fig1}, are swapped following the above mentioned method, while the pair-produced ones remain unchanged so the LCC background can be separately under control. Fig.~\ref{fig:fig4} (b), (c) and (d) show the net electric charge distributions in the transverse plane after incorporating signals of the CME, the CMW and a superposition of both, respectively. It is easy to prove that two kinds of signals are independent, namely, the CME signal has no effect on the CMW observable and vice versa, so the superposition can be decomposed without mutual interaction. Note that the signals in Fig.~\ref{fig:fig4} are intentionally enhanced for better visualization and only one of two dipole and quadrupole configurations is presented as an example. The strengths of both signals, $S_{\rm CME}$ and $S_{\rm CMW}$, are quantified by the number of pairs being interchanged. Our goal is to quantitatively find out the maximum strength of signal that can be tolerated within experimental precision.

The influences of four key parameters (multiplicity, $f_{\rm LCC}$, $S_{\rm CME}$ and $S_{\rm CMW}$) on three observables ($\Delta \delta$, $\Delta \gamma$ and the slope) are summarized in Tab.~\ref{tab:tab2}. First, to accommodate both signals, which increase the $\Delta \gamma$ and the slope, the $f_{\rm LCC}$ must be reduced. However, this will lead to the further decrease of the $\Delta \delta$. Consequently, as a compensation, the multiplicity needs to be reduced as well, which is the only viable solution for making the signal and the background coexist. Note that the multiplicity is a robust observable rather than a freely adjustable parameter but the measurement allows up to 10\% uncertainty around the mean values~\cite{ALICE:2019hno}. Therefore, we reduce the multiplicity by a few percent ($<$10\%) in each centrality, ensuring that the $A_{\rm ch}$ distribution remains comparable to the data. With the fixed multiplicity, the $f_{\rm LCC}$ is then tweaked as small as it can go to match the limits of the measured charge balance function and the $\Delta \delta$. Compared to the values in Tab.~\ref{tab:tab1}, $f_{\rm LCC}$ are diminished by $\sim$15\%,  $\sim$10\% and $\sim$1\% in the most central, the semi-central and the peripheral collisions, respectively.

\begin{table}
\vspace{0.6cm}
\renewcommand\arraystretch{1.6}
\captionsetup{justification=raggedright}
\setlength{\tabcolsep}{4mm}
\caption{The impact of four key parameters on the CME and the CMW observables.}
\begin{tabular}{c || c | c | c }
& $\Delta \delta$ & $\Delta \gamma$ & Slope \\
\midrule[1.5pt]
Mult. $\searrow$ & $\nearrow$ & $\nearrow$ & - \\
\hline
$f_{\rm LCC}$ $\searrow$ & $\searrow$ & $\searrow$ & $\searrow$ \\
\hline
$S_{\rm CME}$ $\nearrow$ & $\searrow$ & $\nearrow$ & - \\
\hline
$S_{\rm CMW}$ $\nearrow$ & - & - & $\nearrow$ \\
\end{tabular}
\label{tab:tab2}
\end{table}

In the absence of the LCC background, i.e., when all particles are single-produced, it is discovered that the natural charge fluctuation can produce a $\Delta \delta$ and a $\Delta \gamma$ on the order of $\sim 10^{-7}$-$10^{-6}$, serving as a baseline. Swapping the charge once to create one pair of dipole moment is able to generate a $\Delta \delta \sim$ -$10^{-5}$ and a $\Delta \gamma \sim$ $10^{-5}$. Since the variation of $\Delta \delta$ and $\Delta \gamma$ is simply proportional to the interchange time, $S_{\rm CME}$, following the rule of $S_{\rm CME}^2$, switching 3 times will almost increase the observables by one order of magnitude. Similarly, creating one quadrupole gives rise to the $\Delta v_2 \sim 10^{-3}$ when $A_{\rm ch} \approx \pm 0.1$, and the further variation follows the rule of 2$S_{\rm CMW}$. These simple rules serve as reference when adding the signal.

\begin{figure}
\centering
\includegraphics[width=\linewidth]{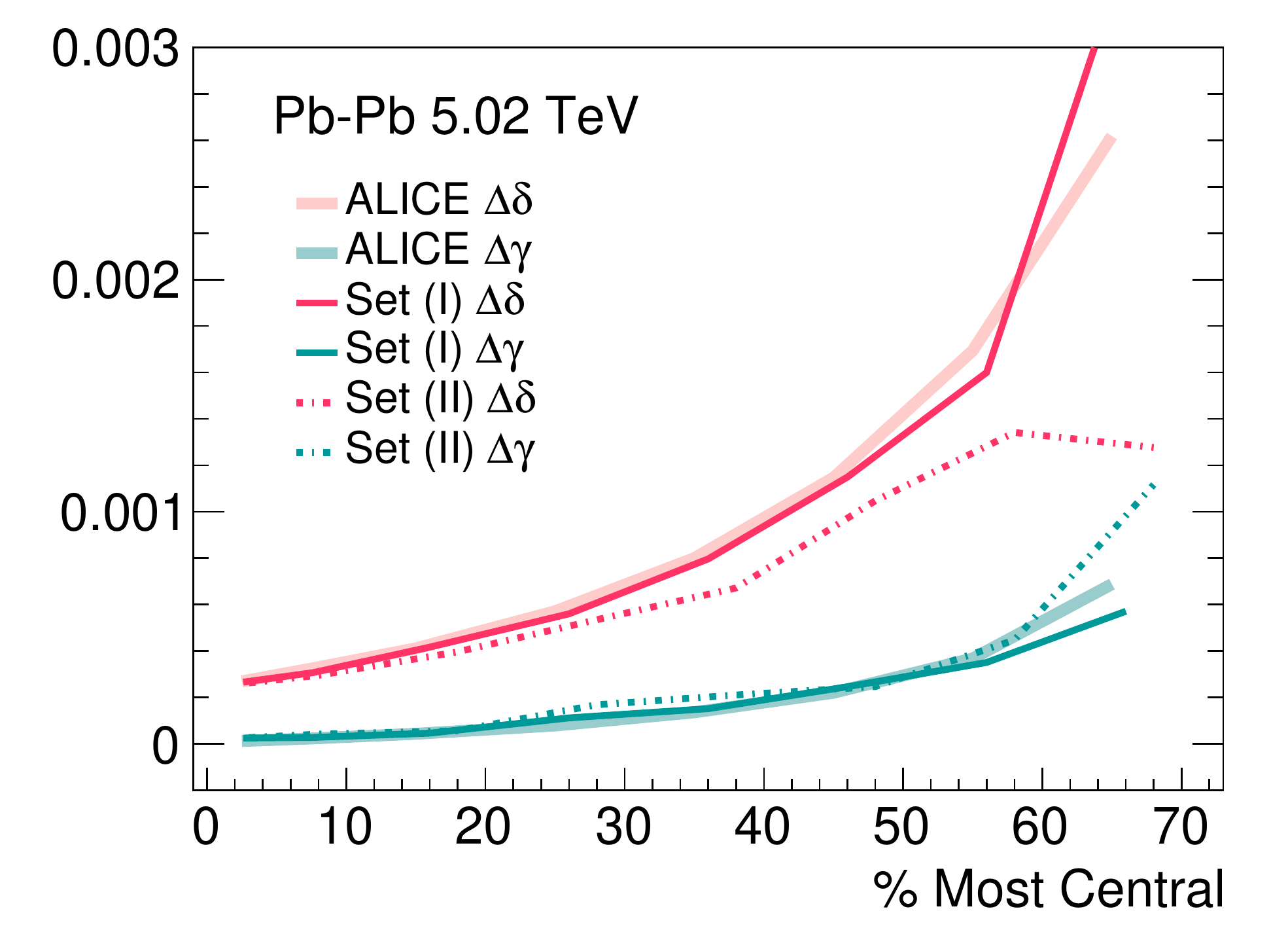}
\captionsetup{justification=raggedright}
\captionof{figure}{The CME observables $\Delta\gamma$ and $\Delta\delta$ as functions of centrality when the signal is imported. Set (I) is still comparable with the ALICE data while set (II) is not, indicating the maximum allowable strength.}
\label{fig:fig5}
\end{figure}

Two sets of CME signal, (I) and (II), are implemented. In set (I), the $S_{\rm CME}$ is linearly related to the multiplicity, so the values vary event-by-event from 1 to 3 for central and semi-central collisions and from 0 to 1 for peripheral collisions. In set (II), the $S_{\rm CME}$ is fixed to 3 (1) for central and semi-central (peripheral) collisions. Fig.~\ref{fig:fig5} shows the comparison between two sets and the ALICE data. It is obvious to see that the finely tuned set (I) is still roughly consistent with the data. However, the fixed set (II), providing signals overmuch, fails, in particular for peripheral collisions, which indicates that the signal does not likely to appear there. The deviation from data continues to increase as the $S_{\rm CME}$ grows. In 30-40\% centrality, the mean $S_{\rm CME}$ value is $\sim$1 and the net CME-induced $\Delta \gamma$ when $S_{\rm CME}$ = 1 is 2$\cdot 10^{-5}$. Compared to the total $\Delta \gamma$ of 1.5$\cdot10^{-4}$, therefore, the allowable maximum fraction of CME signal should be no more than $\sim$13\%. Likewise, the average allowable $S_{\rm CMW}$ in 30-40\% centrality is $\sim$0.5, meaning that one signal is permitted in every two events, and the calculated maximum fraction of CMW in the slope is found to be $\sim$2\%. A larger signal, e.g., $S_{\rm CMW}$ = 1, would result in the increase of the slope by a factor of 3. Note that such estimations do not take any measuring factor into account. Considering various experimental uncertainties (detector-wise and methodology-wise), the final measured signals, even existing, would be further reduced. In a word, any experimental results larger than these upper limits might be unreasonable and there are very few opportunities to capture such tiny signals in LHC energy, matching the current ALICE and CMS conclusions. 

$\it{Conclusion. -}$
In heavy-ion collisions, when studying the charge dependent correlations to search for anomalous chiral effects, the LCC effect is recognized as one of the most important backgrounds. Meanwhile, the CME and the CMW, despite sharing a similar background mechanism, have always been handled as two separate analyses. In this letter, we present a simulation using a modified BW model, which naturally describes most experimental results, including the CME and the CMW observables, at the same time. Such a universal description, for the first time, unifies the CME and the CMW studies and quantitatively reveals the connections between the observable, the signal and the LCC background. Following the $principle~of~parsimony$, we argue that, in LHC energy, the measured results of both the CME and the CMW can be fully interpreted by the LCC entwined with the collective flow. We then propose a phenomenological approach of incorporating the CME and the CMW signals to quantify the maximum allowable strength of the signal. Our calculation shows that the CME and CMW fractions in the observables, even in the most ideal condition, should be no more than 13\% and 2\% respectively. Such a global constraint provides a fresh perspective to understand the experimental data.
 
The BW model is quite straightforward and the realistic environment is obviously more complicated. However, the calculation in this work is tightly based on experimental results and the success of this model, particularly in unifying the CME and the CMW observables, may enlighten other more fundamental studies.
We strongly suggest that this method should be extended to RHIC energy, where signals are expected to be stronger due to possibly longer life time of the magnetic field, to further nail down the interpretation. It is also desirable to introduce identified particles to investigate more subtle correlations. 

We are grateful to our collaborators in the ALICE and the STAR experiments for enlightening discussions and suggestions. We appreciate J. Liao for helpful comments from theoretical standpoint. We also thank W.-B. He and C. Zhong for their effort on maintaining computing resources. This work is supported by the National Key Research and Development Program of China (Nos. 2018YFE0104600, 2016YFE0100900), the National Natural Science Foundation of China (Nos. 11890710, 11890714, 12061141008, 11975078, 11421505), the Strategic Priority Research Program of Chinese Academy of Sciences (No. XDB34030000) and the Guangdong Major Project of Basic and Applied Basic Research (No. 2020B0301030008). Q.-Y. S. is also sponsored by the Shanghai Rising-Star Program (20QA1401500).

\bibliographystyle{utphys}
\bibliography{bibliography}

\end{document}